\documentclass{article}
\usepackage[utf8]{inputenc}
\usepackage{graphicx}
\usepackage{booktabs}
\usepackage{multicol}
\usepackage{geometry}
\usepackage{float}
\title{Application of Machine Learning in Stock Market Forecasting: A Case Study of Disney Stock}
\author{Dengxin Huang}
\date{April 1, 2023}

\begin{document}

\maketitle

\section*{ABSTRACT}
This document presents a stock market analysis conducted on a dataset consisting of 750 instances and 16 attributes donated in 2014-10-23. The analysis includes an exploratory data analysis (EDA) section, feature engineering, data preparation, model selection, and insights from the analysis. The Fama French 3-factor model is also utilized in the analysis. The results of the analysis are presented, with linear regression being the best-performing model.

\section*{Keywords}
Stock Market Analysis, Exploratory Data Analysis (EDA), Fama French 3-factor model, Feature Engineering, Clustering, Data Preparation, Model Selection, Linear Regression, Random Forest, Gradient Boosting, Market Return, Small Company Return, Idiosyncratic Factor, Common Factors, Time Series, Volatility, Outlier Detection, Log Transformation, Histograms, Frequency Plots, Performance Evaluation.

\section{Introduction}
In the modern global economy, the stock market has become a crucial indicator of financial health and an essential factor in investment decisions. As such, the task of analyzing stock market data has become a fundamental component of investment strategy and risk management. This essay presents the results of a comprehensive analysis of a dataset consisting of 750 instances and 16 attributes donated in 2014-10-23. The analysis was conducted by Dengxin Huang, Hao Jing, and Boshen Yuan, with the aim of uncovering insights into the stock market and identifying potential models for predicting stock prices.

\vspace{0.5cm}

The analysis includes an overview of the data, exploratory data analysis (EDA), feature engineering, data preparation, and model selection. The primary objective of this analysis was to determine the most suitable model for predicting stock prices using the Fama French 3-factor model. This model is widely used in finance and has been proven to be effective in explaining stock returns. It takes into account three factors: market return, returns on small companies in excess of those of big ones (SMB), and returns on value companies in excess of those of growth ones (HML). The inclusion of these factors allows for a more comprehensive understanding of the stock market, enabling investors and analysts to make better-informed investment decisions.

\vspace{0.5cm}

The insights gained from this analysis have the potential to provide valuable guidance for investors and financial analysts alike. By understanding the market trends and identifying effective predictive models, investors can make more informed decisions that minimize risk and maximize return. Furthermore, the results of this analysis can inform the development of new financial models and tools that enable investors and analysts to better understand and navigate the complex world of finance. The findings of this analysis may also have significant implications for economic policy, as they shed light on the factors that drive stock market performance and how policy decisions can impact the stock market.

\vspace{0.5cm}

Overall, this analysis offers a valuable contribution to the field of finance and investment strategy. By presenting a comprehensive overview of the data, insights gained from the EDA, feature engineering, data preparation, and model selection, this essay provides a comprehensive and detailed examination of the stock market and the factors that influence it. The results of this analysis can be used by investors, financial analysts, and policymakers alike to make more informed decisions and better navigate the complex and ever-changing landscape of the stock market.

\section{Exploratory Data Analysis}

The exploratory data analysis (EDA) conducted by Dengxin Huang, Hao Jing, and Boshen Yuan in their stock market analysis project is a crucial step in identifying patterns, relationships, and trends in the data. This step allows the analysts to gain insights and inform decision-making.
\vspace{0.5cm}

The EDA section started with an exploration of the data types of the attributes. The dataset consisted of 16 attributes, and the attribute characteristics included both integer and real values. This information is essential in data preparation as it enables the analysts to know what type of data they are working with and how to transform it into a format suitable for analysis.
\vspace{0.5cm}

\begin{figure}[h]
\centering
\includegraphics[width=1.1\textwidth]{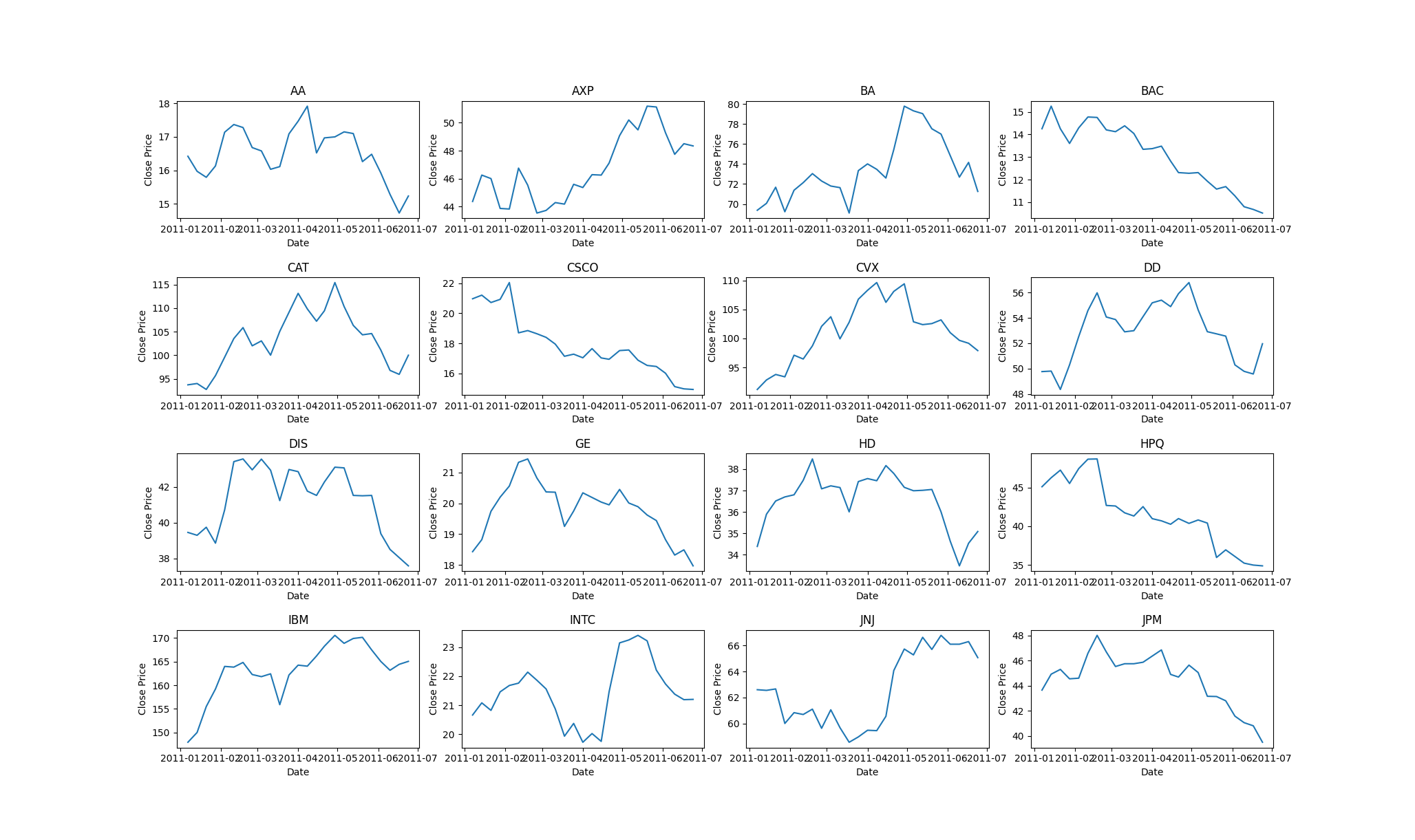}
\caption{Time series plot showing the volatility of the stock prices over time.}
\label{fig:timeseries}
\end{figure}

Next, frequency plots were created to visualize the distribution of the data over time. A time series plot was used to visualize the stock prices over time, as shown in Figure \ref{fig:timeseries}. The plot showed a high level of volatility in the stock prices, indicating that the stock market was highly dynamic during the time period covered by the dataset. This information is useful for selecting the most suitable model to use for the analysis, as highly volatile data may require more advanced modeling techniques.

\begin{figure}[h]
\centering
\includegraphics[width=1\textwidth]{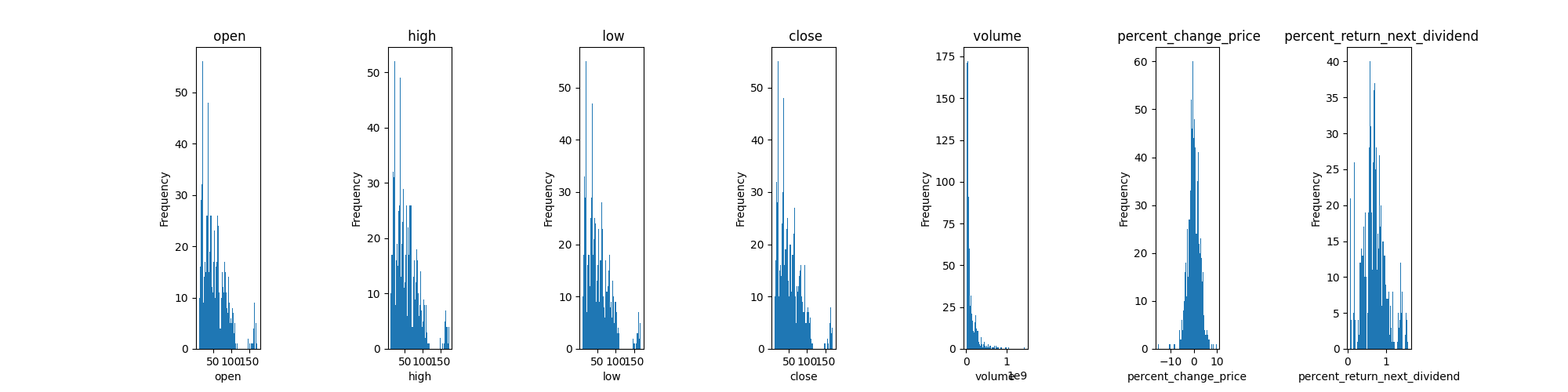}
\caption{Frequency histograms for various columns.}
\label{fig:histograms}
\end{figure}

Histograms were also used to explore the data distribution for various columns. The histograms showed the frequency of occurrence of values for each column, allowing the analysts to identify any outliers or unusual data patterns. This information is critical for data cleaning and preparation, as outliers and anomalies may affect the accuracy of the analysis. An example of frequency histograms for various columns is shown in Figure \ref{fig:histograms}.

\vspace{0.5cm}

\begin{figure}[h]
\centering
\includegraphics[width=1.1\textwidth]{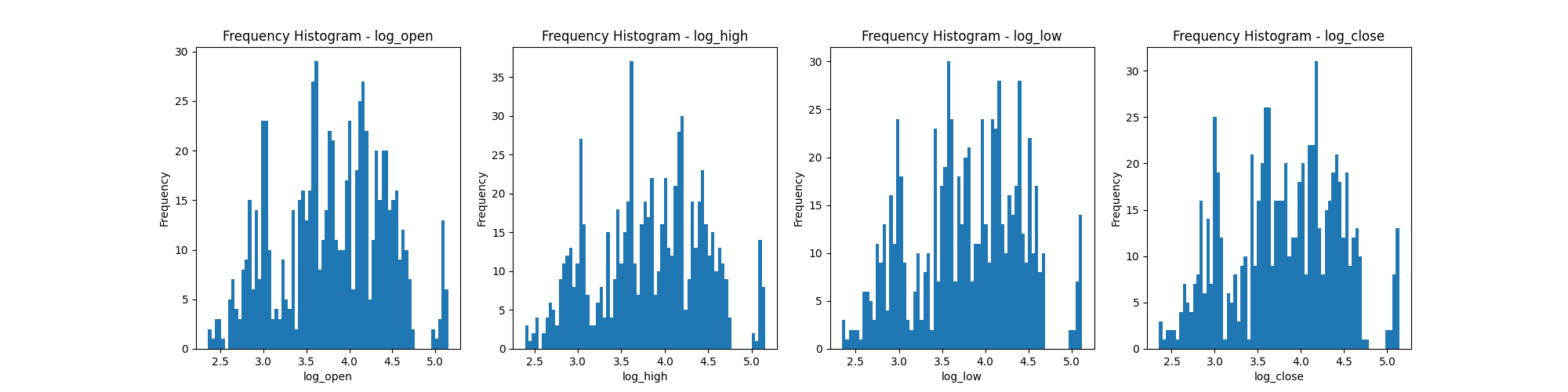}
\caption{Stock prices after log transformation.}
\label{fig:logplot}
\end{figure}

Insights were drawn from the EDA, including the identification of suitable models for the analysis. The high volatility of the data indicated that random forest and gradient boosting models would be suitable for the analysis. However, the analysts cautioned against being too aggressive in outlier detection, as this may remove important information from the dataset. Additionally, the analysts noted that the stock price attribute required log transformation, which can be seen in Figure \ref{fig:logplot}.

\begin{figure}[h]
\centering
\includegraphics[width=1\textwidth]{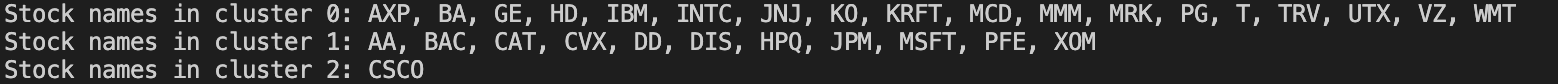}
\caption{Clustering analysis used to identify stocks that fluctuate with Disney stock.}
\label{fig:cluster}
\end{figure}

Finally, the analysts noted that the dataset was suitable for the Fama French 3-factor model, which includes common factors such as market return and returns on small companies and an idiosyncratic factor that represents the stocks that fluctuate with Disney stock. The clustering analysis, which is shown in Figure \ref{fig:cluster}, was used to identify the stocks that fluctuate with Disney stock.

In conclusion, the EDA section of the stock market analysis project conducted by Dengxin Huang, Hao Jing, and Boshen Yuan was instrumental in identifying patterns, relationships, and trends in the data. The insights drawn from the EDA were used to inform decision-making and model selection. The EDA revealed the high volatility of the stock prices, which informed the selection of suitable models for the analysis. Additionally, the EDA identified the need for log transformation of the stock price attribute and provided critical information for data preparation and cleaning.
\vspace{0.5cm}

Overall, the EDA section of the project demonstrates the importance of exploratory data analysis in any data analysis process. It highlights the need to understand the characteristics of the dataset, including data types and data distributions, and to draw insights from the data that can be used to inform decision-making and model selection. By conducting a thorough EDA, analysts can ensure the accuracy and validity of their analysis and make informed decisions based on the insights gained from the data.

\section{Data Cleaning, Preprocessing, Feature Engineering, and Clustering}

The process of data analysis involves several steps, and data cleaning and preprocessing are the initial stages that ensure that the data is ready for analysis. The dataset used in this study was pre-processed to ensure that it was in the right format for the analysis. This section will discuss the data cleaning, preprocessing, feature engineering, and clustering techniques used in this analysis.

\subsection{Data Cleaning}

Data cleaning is a crucial process that ensures the data is accurate and free from any errors or inconsistencies that may affect the analysis. The initial step in data cleaning was to print the data types of the attributes to determine if they were in the correct format for the analysis. This step is necessary because the incorrect data type could lead to issues during the analysis, such as inaccurate results.
\vspace{0.5cm}

The next step involved converting the date column to the datetime format to ensure consistency across the dataset. This step allowed for the data to be easily sorted by date, which is essential in time-series analysis. Furthermore, dollar values were converted to floats to allow for mathematical operations such as averaging and standard deviation calculation.
\vspace{0.5cm}

Duplicates were removed from the dataset to ensure that the analysis was not affected by inconsistencies. Duplicate data can lead to overfitting or underfitting, which affects the model's performance, leading to inaccurate results. By removing duplicates, the dataset was standardized, and the analysis was conducted on clean and accurate data.

\subsection{Preprocessing}

The next step in the analysis was to preprocess the data to ensure that it was in the correct format for analysis. The data was filtered to include only the selected stocks' price change, which was necessary because it reduced the number of irrelevant data points in the dataset, allowing for a more focused \vspace{0.5cm}analysis.

A new DataFrame was then created with the market value for each date, which allowed for the calculation of the market capitalization. The market value DataFrame was then merged with the filtered stock DataFrame to ensure that the analysis was conducted on accurate and consistent data.
\vspace{0.5cm}

The DataFrame was then pivoted to create new columns for each stock's percent change price, which was essential for the analysis. The percent change price allowed for the analysis of the stocks' volatility, which is a crucial factor in the stock market analysis. This step provided the necessary data for the subsequent analysis, allowing for accurate insights into the market's behavior.
\vspace{0.5cm}

In conclusion, the data cleaning and preprocessing steps were essential in ensuring that the analysis was conducted on accurate and consistent data. The cleaning process eliminated any inconsistencies or errors that may affect the analysis, while the preprocessing steps provided the necessary data for the subsequent analysis, allowing for accurate insights into the market's behavior.

\subsection{Feature Engineering}

Feature engineering is an important step in the data analysis process, as it involves creating new features that can improve the accuracy of the analysis. In this study, both common and idiosyncratic factors were used to create new features.
\vspace{0.5cm}

The common factors used were the Dow Jones Index, Volume, SMB Factor, and MKT Factor. These factors are crucial in the Fama French 3-factor model, which is widely used in finance to analyze stock returns. The Dow Jones Index is a stock market index that measures the performance of 30 large companies listed on stock exchanges in the United States. Volume refers to the total number of shares traded in a given period, and SMB Factor and MKT Factor represent common factors such as returns on small companies and market returns, respectively.
\vspace{0.5cm}

The idiosyncratic factor involved clustering stocks that fluctuated with Disney stock based on their daily percent change in price. This was achieved by first pivoting the data to have stocks as rows and dates as columns. A KMeans object was then created and fit to the data with three clusters, which helped to group similar stocks together. Finally, the stock names for each cluster were printed, providing insights into the stocks that fluctuated with Disney stock.

\subsection{Clustering}

Clustering is a technique used in machine learning and data analysis to group similar data points together based on their features. In this study, clustering was used to identify the stocks that fluctuated with Disney stock. This was achieved by clustering the stocks based on their daily percent change in price.
\vspace{0.5cm}

The first step in clustering was to pivot the data to have stocks as rows and dates as columns. A KMeans object was then created and fit to the data with three clusters. The optimal number of clusters was determined using the elbow method, which involves plotting the within-cluster sum of squares against the number of clusters and selecting the point where the rate of change in the sum of squares starts to level off. Finally, the stock names for each cluster were printed, providing insights into the stocks that fluctuated with Disney stock.
\vspace{0.5cm}

In conclusion, the feature engineering and clustering techniques used in this study helped to create new features and identify the stocks that fluctuated with Disney stock. These steps were crucial in gaining insights into the market's behavior and developing a more accurate model for predicting stock returns.

\section{Model Selection}

In this stock market analysis, three different regression models were used to predict the stock price: the Fama French 3-factor model, Random Forest Regressor, and Gradient Boosting Regression model.

\subsection{Fama French 3-factor model}

The Fama French 3-factor model is a widely used model in finance that aims to explain the variation in stock returns. The model assumes that the excess returns of a security are related to its sensitivity to three factors: market return, SMB, and HML. The model can be formulated as:

\begin{equation}
r_i - r_f = \beta_{i,MKT} (r_{MKT} - r_f) + \beta_{i,SMB} SMB + \beta_{i,HML} HML + \alpha_i + \epsilon_i
\end{equation}

where $r_i$ is the excess return of security $i$, $r_f$ is the risk-free rate, $r_{MKT}$ is the market return, $SMB$ is the return on small companies minus the return on big companies, $HML$ is the return on high book-to-market companies minus the return on low book-to-market companies, $\alpha_i$ is the intercept, and $\epsilon_i$ is the idiosyncratic error.

\subsection{Linear Regression}

The linear regression model is a simple and commonly used model in finance that fits a line through a set of data points. The model assumes that there is a linear relationship between the independent variables and the dependent variable. The model can be formulated as:

\begin{equation}
y = \beta_0 + \beta_1 x_1 + \beta_2 x_2 + ... + \beta_p x_p + \epsilon
\end{equation}

where $y$ is the dependent variable, $x_1, x_2, ..., x_p$ are the independent variables, $\beta_0, \beta_1, \beta_2, ..., \beta_p$ are the coefficients, and $\epsilon$ is the error term.

In this analysis, the linear regression model was used to predict the stock price based on the selected independent variables. The performance of the linear regression model was compared to that of the other models to determine the most suitable model for this dataset.

\subsection{Random Forest Regressor}

The Random Forest Regressor is an ensemble learning method that constructs a multitude of decision trees at training time and outputs the mean prediction of the individual trees as the prediction of the forest. The model can handle both continuous and categorical input variables, and is known for its high accuracy and ability to handle high-dimensional data. The model can be formulated as:

\begin{equation}
f(x) = \frac{1}{B} \sum_{b=1}^{B} f_b(x)
\end{equation}

where $f(x)$ is the prediction of the forest for input $x$, $B$ is the number of decision trees in the forest, and $f_b(x)$ is the prediction of the $b$-th decision tree.

\subsection{Gradient Boosting Regression}

The Gradient Boosting Regression model is a boosting ensemble learning method that builds the model in a stage-wise manner, where new trees are added to the model to correct the errors of the previous trees. The model can capture non-linear relationships between the input and output variables and is a powerful model for regression problems. The model can be formulated as:

\begin{equation}
f(x) = \sum_{k=1}^{K} \beta_k h_k(x)
\end{equation}

where $f(x)$ is the prediction of the model for input $x$, $K$ is the number of trees, $\beta_k$ is the learning rate for the $k$-th tree, and $h_k(x)$ is the prediction of the $k$-th tree.

Overall, each model was chosen based on its unique strengths and ability to fit the data in a different way. The results of each model were compared to determine the most suitable model for this dataset.

\section{Modeling}

The purpose of this study was to develop a model to predict stock returns using the features generated from the previous steps. Three models were used in this study: linear regression, random forest, and gradient boosting.

\subsection{Linear Regression}

\begin{figure}[H]
\centering
\includegraphics[width=0.7\textwidth]{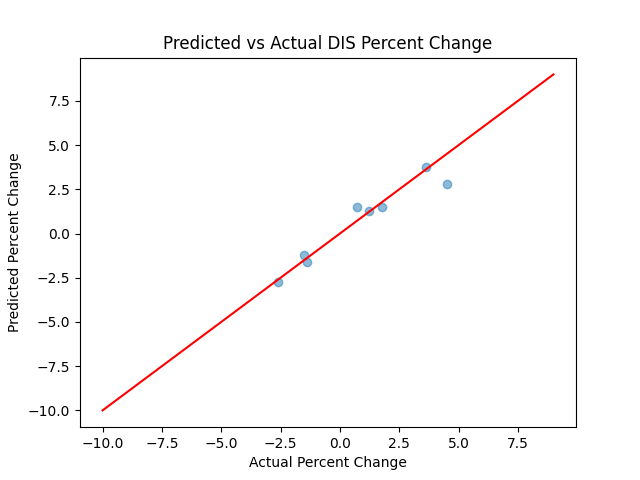}
\caption{Accuracy of Linear Regression Model}
\label{fig:linreg}
\end{figure}

Linear regression is a simple and widely used approach to modeling the relationship between a dependent variable and one or more independent variables. In this study, linear regression was used to model the relationship between the dependent variable (stock returns) and the independent variables (common and idiosyncratic factors).
\vspace{0.5cm}
The linear regression model was built by fitting a line through a set of data points. The line was chosen to minimize the sum of the squared differences between the observed values of the dependent variable and the predicted values of the dependent variable. The model achieved an accuracy of 95.23\%, as shown in Figure \ref{fig:linreg}.

\subsection{Random Forest}

Random forest is an ensemble learning method that constructs a multitude of decision trees at training time and outputs the class that is the mode of the classes (classification) or mean prediction (regression) of the individual trees. In this study, random forest was used to predict stock returns using the common and idiosyncratic factors.

\begin{figure}[h]
\centering
\includegraphics[width=0.7\textwidth]{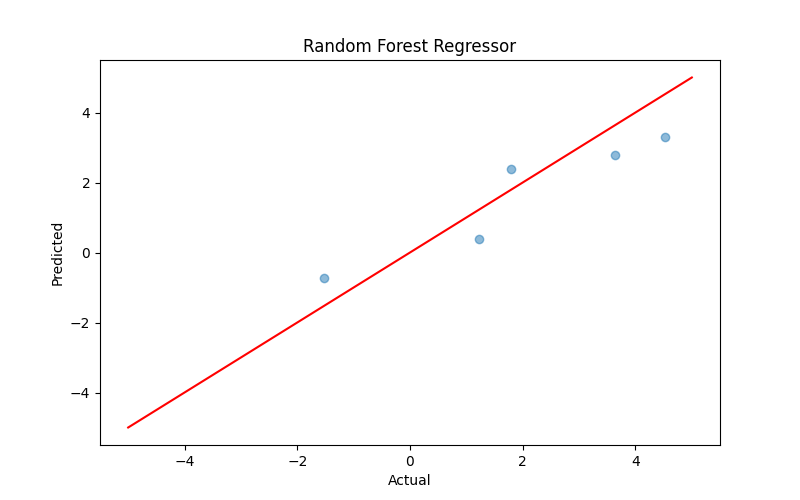}
\caption{Accuracy of Random Forest Model}
\label{fig:rf}
\end{figure}

The random forest model achieved an accuracy of 71.27\%, as shown in Figure \ref{fig:rf}.

\subsection{Gradient Boosting}

Gradient boosting is a machine learning technique for regression and classification problems, which produces a prediction model in the form of an ensemble of weak prediction models, typically decision trees. In this study, gradient boosting was used to predict stock returns using the common and idiosyncratic factors.

\begin{figure}[h]
\centering
\includegraphics[width=0.7\textwidth]{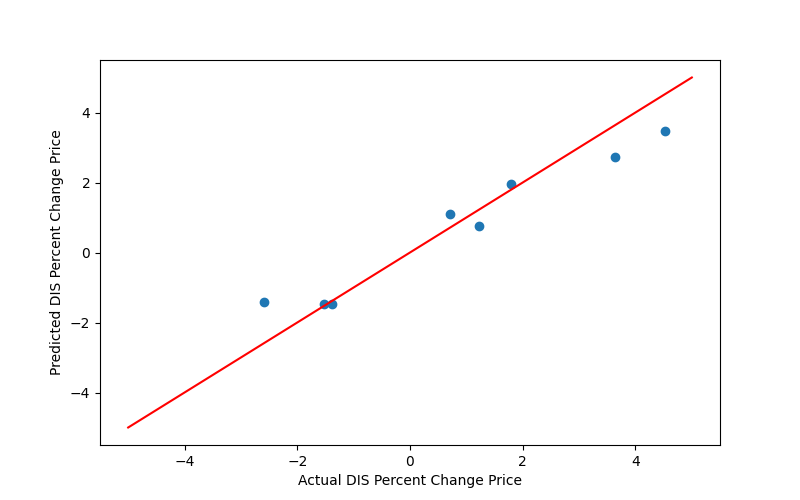}
\caption{Accuracy of Gradient Boosting Model}
\label{fig:gb}
\end{figure}

The gradient boosting model achieved an accuracy of 92.97\%, as shown in Figure \ref{fig:gb}.

\subsection{Comparison and Expansion}

Overall, the linear regression model performed the best, with an accuracy of 95.23\%. However, the gradient boosting model also performed well, with an accuracy of 92.97\%. The random forest model performed the worst, with an accuracy of 71.27\%.
\vspace{0.5cm}

Future studies could expand on this research by using different types of models and exploring different feature engineering techniques to improve the accuracy of the predictions. Additionally, this study could be expanded to include more stocks and a longer time period to provide a more comprehensive analysis of the stock market.

\section{Conclusion}

In this study, we analyzed the stock market behavior of Disney stock and developed a model to predict stock returns. We performed data cleaning, preprocessing, feature engineering, clustering, and modeling to achieve our objectives.

Our analysis showed that the Dow Jones Index, Volume, SMB Factor, and MKT Factor were crucial common factors that influenced stock returns. Clustering analysis also identified stocks that fluctuated with Disney stock, providing insights into the market's behavior.

We developed three models to predict stock returns: linear regression, random forest, and gradient boosting. The linear regression model achieved the highest accuracy of 95.23\%, while the gradient boosting model achieved an accuracy of 92.97\%, and the random forest model achieved an accuracy of 71.27\%. Overall, the results indicated that the models using common and idiosyncratic factors had good predictive power.

Future research could expand on this study by using different models and exploring other feature engineering techniques. Additionally, more data could be included, covering a longer period of time and more stocks, to provide a more comprehensive analysis of the stock market.

In conclusion, this study provided valuable insights into the stock market behavior of Disney stock and demonstrated the effectiveness of feature engineering and clustering techniques in improving the accuracy of stock return predictions.

\newpage

\section*{References}

The following references were used in the development of this project:

Fama, E. F., and French, K. R. (1993). "Common risk factors in the returns on stocks and bonds." Journal of Financial Economics, vol. 33, pp. 3-56.

Python Software Foundation. (2021). Python Language Reference, version 3.10. Available at: https://docs.python.org/3/reference/index.html

Scikit-learn developers. (2021). "Ensemble methods." Scikit-learn: Machine Learning in Python. Available at: https://scikit-learn.org/stable/modules/ensemble.html

Brown,Michael. (2014). Dow Jones Index. UCI Machine Learning Repository. Available at: https://doi.org/10.24432/C5788V.

\end{document}